\documentclass[twocolumn,prb,amsmath,amssymb,floatfix,superscriptaddress]{revtex4-2}
\usepackage{amsmath}
\usepackage{amssymb}
\usepackage{graphicx}
\usepackage{bm}
\usepackage{color}
\usepackage{hyperref}
\usepackage{graphicx}
\usepackage{upgreek}
\usepackage{scalerel}
\usepackage{multirow}
\usepackage{pgffor}
\usepackage{pdfpages}
\usepackage{ mathdots}
\usepackage{float}
\usepackage{physics} 
\usepackage{lscape}   
\usepackage{color, colortbl}
\usepackage[table]{xcolor}
\usepackage{comment}  
\makeatletter
\AtBeginDocument{\let\LS@rot\@undefined}
\makeatother

\begin{document}

\title{Floquet engineering bulk odd-frequency superconducting pairs}
\author{Jorge Cayao}
\affiliation{Department of Physics and Astronomy, Uppsala University, Box 516, S-751 20 Uppsala, Sweden}
 \author{Christopher Triola} 
\affiliation{Los Alamos National Laboratory, Los Alamos, New Mexico 87544, USA}
\affiliation{Department of Physics and Astronomy, Uppsala University, Box 516, S-751 20 Uppsala, Sweden}
\author{Annica M. Black-Schaffer}
\affiliation{Department of Physics and Astronomy, Uppsala University, Box 516, S-751 20 Uppsala, Sweden}

\date{\today}
\begin{abstract}
We introduce the concept of Floquet odd-frequency superconducting pairs and  establish their emergence in time-periodic conventional superconductors.  We show that these exotic Cooper pairs are possible because the Floquet modes in time-periodic systems  provide an additional index (a Floquet index) that broadens the classification of superconducting pair symmetries, with no analog in the static regime. Our results thus put forward a different route for odd-frequency superconducting pairs and pave the way for Floquet engineered dynamical superconducting states. 
\end{abstract}\maketitle

\section{Introduction}
Since its discovery, superconductivity has  garnered widespread attention not only owing to its  fundamental mechanisms but also due to its large number of applications, making it one of the core areas in condensed matter. Recent developments have reported a plethora of remarkable superconducting states, such as chiral superconductors \cite{kallin2016chiral}, high-temperature superconductors \cite{drozdov2019superconductivity}, topological superconductors \cite{Sato_2017}, superconducting metamaterials \cite{singh2014nat}, magic angle superconductors \cite{cao2018unconventional}, and nematic superconductors \cite{yonezawa2017thermodynamic}. This vast diversity of superconductors is highly reliant on the symmetries of their fundamental building blocks, the Cooper pairs.

The symmetries of Cooper pairs are constrained by the fermionic nature of the constituent electrons \cite{RevModPhys.63.239}, which  imposes antisymmetry on the Cooper pair amplitude under the interchange of all quantum numbers  of the paired electrons, including the exchange of their time coordinates. This allows for the usual pair correlations between electronic states at equal times but also permits electron pairing at different times. Remarkably, such unequal time pairing  enables the emergence of Cooper pairs with a pair amplitude that is odd under the exchange of time coordinates, or, equivalently, odd in frequency \cite{bere74,PhysRevLett.66.1533,PhysRevB.45.13125,PhysRevLett.86.4096,Kadigrobov01,PhysRevB.76.054522,PhysRevLett.98.037003,Eschrig2007}. Since  odd-frequency (odd-$\omega$) pair amplitudes  vanish at equal times,  this type of pairing is intrinsically non-local in time and represents a truly dynamical effect. 

The emergence of odd-$\omega$ pairs has been shown to be related to the breaking of system symmetries and a variety of superconducting systems are believed to host these exotic correlations  \cite{RevModPhys.77.1321,Nagaosa12, Balatsky2017, cayao2019odd, triola2020role}. In particular, superconducting heterostructures \cite{PhysRevLett.86.4096,Kadigrobov01,PhysRevB.76.054522,PhysRevLett.98.037003,Eschrig2007,PhysRevB.92.100507,PhysRevB.96.155426,PhysRevB.97.134523,PhysRevB.100.104511,PhysRevB.101.195303}, with experimental observations in ferromagnetic junctions \cite{zhu2010angular,di2015signature,PhysRevX.5.041021}, and  superconductors with multiple degrees of freedom \cite{PhysRevB.88.104514,PhysRevB.90.184517,
PhysRevB.90.220501,PhysRevB.92.094517,PhysRevB.93.201402,10.1093/ptep/ptw094,PhysRevB.101.180512} are the most notable examples. In heterostructures, odd-$\omega$ pairs are induced by the breaking of spatial translation symmetry, while in multiband superconductors, hybridization between bands, which can be seen as an intrinsic symmetry breaking effect, gives rise to odd-$\omega$ pairs. 

Despite the dynamical nature of odd-$\omega$ pairing, the overwhelming majority of previous work has focused on static systems and static properties of odd-$\omega$ pairing. Moreover, the few works which have found odd-$\omega$ pairing induced by time-dependent drives have shown that mere time-dependence is not sufficient to generate odd-$\omega$ pairing, the drive must either break translation invariance \cite{PhysRevB.94.094518} or couple non-trivially to a band index \cite{PhysRevB.95.224518}.   While time-dependent  drives may reduce the order parameter \cite{PhysRevB.76.224522,PhysRevB.90.014515}, e.g.~via heating, there are strategies to mitigate this effect, for example, choosing a drive frequency that is large compared with the relevant energy scales in the system \cite{rudner2020band}. Additionally, external  drives have even been shown to enhance or induce superconductivity \cite{PhysRev.174.482,doi:10.1063/1.107069,fausti2011light,PhysRevB.93.144506,PhysRevB.94.214504,PhysRevLett.120.246402,PhysRevB.100.024513,mitrano2016giant}.

\begin{figure}[!t]
	\centering
	\includegraphics[width=.48\textwidth]{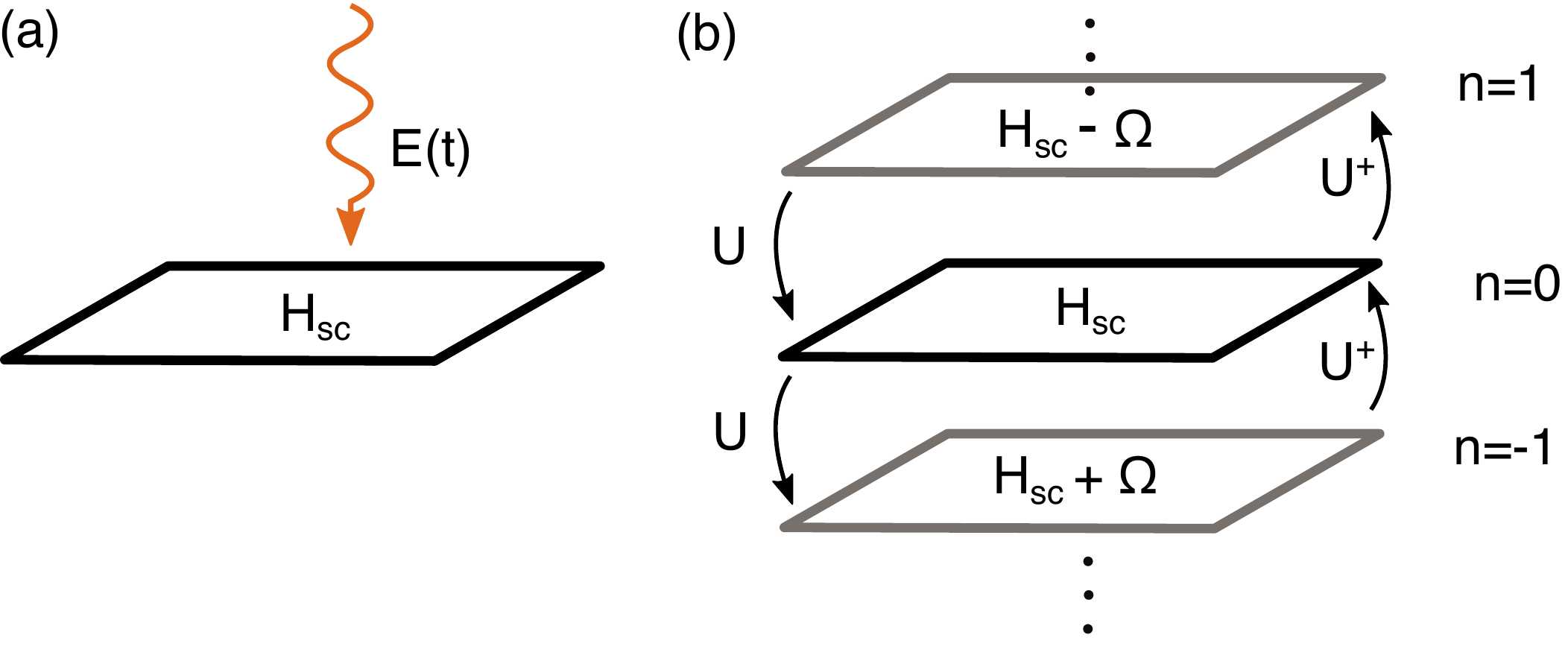}
	\caption{(a) Sketch of superconductor, described by a static Hamiltonian $H_{\rm sc}$ driven by a  time-periodic field $E(t)$ with frequency $\Omega$ (wiggle orange arrow). (b) Within Floquet theory the original problem is described in terms of Floquet bands, or sidebands, labeled by the Floquet index $n$, which are replicas of the undriven system $H_{\rm sc}$ shifted by $n\Omega$  and coupled by $U$ (arc black arrows), which depend on $E(t)$.}
	\label{Fig1}
\end{figure}

In this work we show that odd-$\omega$ superconducting pairs are generically present in time-periodic superconductors, where  properties are described in terms of Floquet bands (see Fig.~\ref{Fig1}). We find that these Floquet bands provide an additional index that, remarkably, broadens the classification of  pair symmetries with no analog in the static regime.  In particular, we introduce the concept of Floquet odd-$\omega$ pairing and demonstrate its emergence in conventional  (spin-singlet,  $s$-wave) superconductors  driven by circularly-polarized light, where it  acquires a large magnitude without any fine-tuning of model parameters and can be controlled by the drive.  The prospect of inducing and controlling dynamical pairs by time-periodic fields opens the route for Floquet engineered dynamical superconducting correlations.

 \section{Floquet pair amplitudes}
To show the emergence of Floquet  odd-$\omega$ Cooper pairs, we first provide a general characterization of their symmetries. To begin, we note that the pair amplitudes of any superconductor (driven or static)  are given by the  anomalous Green's function \cite{zagoskin1998quantum,mahan2013many} $F^{\sigma_{1},\sigma_{2}}(\boldsymbol{k}_{1},\boldsymbol{k}_{2};t_{1},t_{2})=-i\expval{\mathcal{T}c_{\boldsymbol{k}_{1},\sigma_{1}}(t_{1})c_{\boldsymbol{k}_{2},\sigma_{2}}(t_{2})}$, where $\mathcal{T}$ is the time-ordering operator and $c_{\boldsymbol{k},\sigma}(t)$ annihilates an electronic state with spin $\sigma$, momentum $\boldsymbol{k}$, at time $t$. The fermionic nature of electrons then imposes the \textit{antisymmetry condition}  $F^{\sigma_{1},\sigma_{2}}(\boldsymbol{k}_{1},\boldsymbol{k}_{2};t_{1},t_{2})=-F^{\sigma_{2},\sigma_{1}}(\boldsymbol{k}_{2},\boldsymbol{k}_{1};t_{2},t_{1})$, which is responsible for  the different pair symmetries \cite{Balatsky2017}, and, in particular, allows for pair amplitudes that are odd in $\omega$ when $F^{\sigma_{1},\sigma_{2}}(\boldsymbol{k}_1,\boldsymbol{k}_2;t_{1},t_{2})=-F^{\sigma_{1},\sigma_{2}}(\boldsymbol{k}_1,\boldsymbol{k}_2;t_{2},t_{1})$. 

While the antisymmetry condition  applies to any two-time pair amplitude \cite{Balatsky2017}, we wish to investigate the properties of pair amplitudes in Floquet systems.
Therefore, we next consider a superconductor in a time-periodic field $E(t)$ with period $T=2\pi/\Omega$, as  shown in Fig.\,\ref{Fig1}(a). This  system is described by  a time-dependent Hamiltonian $H(t)$ that inherits the time periodicity of $E(t)$, such that $H(t)=H(t+T)$. For such a time-periodic Hamiltonian,  the Floquet theorem \cite{ASENS_1883_2_12__47_0,PhysRev.138.B979,PhysRevA.7.2203} allows us to decompose the solutions to the Schr\"{o}dinger equation in terms of harmonics of the fundamental drive frequency $\Omega$. Similarly, we decompose the anomalous Green's function as \cite{RevModPhys.86.779},
\begin{equation}
\label{floquetF}
\begin{split}
&F^{\sigma_{1},\sigma_{2}}(\boldsymbol{k}_{1},\boldsymbol{k}_2;t_{1},t_{2})=\\
&\sum_{m,n}\int \frac{d\omega}{2\pi} {\rm e}^{-i(\omega +n\Omega)t_{1}+i(\omega +m\Omega)t_{2} } F^{\sigma_{1},\sigma_{2}}_{n,m}(\boldsymbol{k}_{1},\boldsymbol{k}_{2},\omega)\,,
\end{split}
\end{equation}
where the coefficients $F_{n,m}$ represent the Floquet pair amplitudes, labeled by the Floquet indices $n,m\in\mathbb{Z}$, and $\omega\in[-\Omega/2,\Omega/2]$. For details see Supplementary Material (SM) \cite{SM}. Noting that the symmetries of the quantity on the left-hand side of Eq. (\ref{floquetF}) are constrained by the fermonic antisymmetry condition, we obtain a constraint for the symmetries of the Floquet modes, $F_{n,m}$,
\begin{equation}
\label{AntisymmetryFloquet}
F^{\sigma_{1},\sigma_{2}}_{n,m}(\boldsymbol{k}_1,\boldsymbol{k}_2;\omega)=-F^{\sigma_{2},\sigma_{1}}_{-m,-n}(\boldsymbol{k}_2,\boldsymbol{k}_1;-\omega)\,.
\end{equation}
Here, the exchange in Floquet indices, $(n,m)\rightarrow(-m,-n)$, stems from the  Floquet decomposition in Eq.\,(\ref{floquetF}), intrinsic in two-time periodic functions \cite{RevModPhys.86.779}. 

 From the constraint in Eq.\,(\ref{AntisymmetryFloquet}) we get all possible Floquet pair symmetries that can emerge in a time-periodic superconducting system, as enumerated in Table \ref{Table1}. Remarkably, there are four different classes of odd-$\omega$ pairs, determined by a combination of the Floquet index, spin (singlet, triplet), and momentum ($s$-wave, $p$-wave etc.). 
 At first sight, it might appear that the Floquet indices $(n,m)$  simply act as a new kind of band index, and thus Eq.~(\ref{AntisymmetryFloquet}) is just a generalization of the antisymmetry constraint for a multiband superconductor \cite{triola2020role} with an arbitrarily large number of band degrees of freedom. To show that this is, in fact, not the case, we note that the antisymmetry condition for a two-band superconductor  is given by $F^{\sigma_{1},\sigma_{2}}_{\alpha,\beta}(\boldsymbol{k}_1,\boldsymbol{k}_2;\omega)=-F^{\sigma_{2},\sigma_{1}}_{\beta,\alpha}(\boldsymbol{k}_2,\boldsymbol{k}_1;-\omega)$ where $\alpha$ and $\beta$ are the band indices \cite{triola2020role}. Thus, the Floquet symmetry constraint, $(n,m)\rightarrow (-m,-n)$, is very different from that of a multiband superconductor. Crucially, this implies that the Floquet pair amplitudes $F_{n,m}$ can, in general, develop  \textit{even} and \textit{odd} terms in the Floquet index which can both exhibit even- or odd-$\omega$ dependence. Yet another difference with the multiband case is that  $n,m\in\mathbb{Z}$, implying that $n,m$ can also be negative. Here, inter-Floquet-band amplitudes $F_{n,-n}$ must be \textit{even} in the Floquet index, while intra-Floquet-band amplitudes $F_{n,n}$ can develop both even and odd terms in the Floquet index, for $n\neq 0$, which, remarkably, can be both even- or odd-$\omega$. This is in stark contrast to the case of multiband systems in which only the interband pairing can be odd in the band index and thus odd in frequency,  for even-parity, spin-singlet superconductors. Hence, the Floquet pair amplitudes are unique and exhibit symmetry classes with no analog in undriven systems.

\begin{table}[t!]
\centering
\begin{tabular}{ |c|c|c|c|c|  }
 \hline
\multirow{2}{*}{\rotatebox[origin=c]{90}{class}}& {\bf Floquet index}& {\bf  Frequency}& {\bf Spin}& {\bf Momentum} \\[0.5ex] 
 & $(n,m)\rightarrow(-m,-n)$ & $\omega\rightarrow-\omega$& $s=\pm1$  & $\boldsymbol{k}_{1}\rightarrow\boldsymbol{k}_{2}$\\
  \hline
1& Even&Even&Singlet&Even
\\
  \hline
2& Even &Even&Triplet&Odd\\
  \hline
3& Odd &Even&Singlet&Odd\\
  \hline
4& Odd &Even&Triplet&Even\\
 \hline
5& Even & Odd&Triplet&Even
\\
  \hline
6& Even &Odd&Singlet&Odd\\
  \hline
7& Odd &Odd&Triplet&Odd\\
  \hline
8& Odd &Odd&Singlet&Even\\
 \hline
\end{tabular}
\caption{All possible symmetries for Floquet pair amplitudes allowed by the antisymmetry condition Eq.\,(\ref{AntisymmetryFloquet}).}
\label{Table1}
\end{table}

\section{Realization of Floquet odd-$\omega$ pairs}
Next we show the emergence of the Floquet pair amplitudes $F_{n,m}$ discussed above. We consider a simple model that both captures the essential physics of the problem and is easy to implement experimentally: an electronic system possessing a spin-singlet $s$-wave superconducting order parameter $\Delta$ driven by a circularly-polarized light field ${\bf E}(t)$  \cite{Wang453,mciver2020light}.  Superconductivity could be realized using a conventional superconductor, or could be engineered using proximity effects in heterostructures for additional control of $\Delta$ \cite{PhysRevB.93.155402,kjaergaard2016quantized}. The superconductor is modeled by $H_{\rm sc}=\xi_{\boldsymbol{k}}\,\tau_{z}+\Delta\,\tau_{x}$ in Nambu space $\Psi_{\boldsymbol{k}}(t)=(c_{{\boldsymbol{k}},\uparrow}(t),c_{-{\boldsymbol{k}},\downarrow}^{\dagger}(t))^{\rm T}$, where  $\xi_{\boldsymbol{k}}={\boldsymbol{k}}^{2}/2m-\mu$ is the kinetic energy with $\boldsymbol{k}=(k_{x},k_{y},k_{z})$, $m$ effective mass,  $\mu$ chemical potential, $\Delta$  considered  to be an input parameter, and  $\tau_{i}$  the $i$-Pauli matrix in Nambu space. The effect  of ${\bf E}(t)$  is taken into account via the minimal coupling ${\boldsymbol{k}}\rightarrow {\boldsymbol{k}}+e\textbf{A}(t)$, where $\textbf{A}(t)$ is the vector potential, ${\bf E}(t)=-\partial_{t}\textbf{A}(t)$, and  $e>0$ the elementary charge. We take $\textbf{A}(t)=A_{0}(-\sin\Omega t,\sigma\cos\Omega t,0)$ with period $T=2\pi/\Omega$, where $\sigma=\pm 1$ denotes left-/right-handed polarizations and $\Omega$ the frequency of the light. Then, if we redefine $\mu$ as $\mu\rightarrow \mu-e^{2}A_{0}^{2}/(2m)$, the time-dependent  Hamiltonian  takes the form $H_{\boldsymbol{k}}(t)=H_{\rm sc}+V_{\boldsymbol{k}}(t)$  with $V_{\boldsymbol{k}}(t)=(e/m){\bf A}(t)\cdot {\boldsymbol{k}} \,\tau_{0}$ a periodic function in $t$. 

We are interested in the pair amplitudes of the time-periodic superconducting system $H_{\boldsymbol{k}}(t)$. 
These are the electron-hole component of the Nambu space Green's function $\hat{\mathcal{G}}(\boldsymbol{k};t_1,t_2)= -i\langle \mathcal{T} \Psi_{\boldsymbol{k}}(t_1) \Psi^\dagger_{\boldsymbol{k}}(t_2) \rangle$, obtained by solving the equation of motion $[i\partial_{t_1} - H_{\boldsymbol{k}}(t_1)]\hat{\mathcal{G}}(\boldsymbol{k};t_1,t_2)=\delta(t_1-t_2)$. Given the time-periodicity of the Hamiltonian $H_{\boldsymbol{k}}(t)$, we can decompose $\hat{\mathcal{G}}$  in terms of Floquet modes \cite{ASENS_1883_2_12__47_0,PhysRev.138.B979,PhysRevA.7.2203} and write the equation of motion for $\hat{\mathcal{G}}$ as \cite{RevModPhys.86.779}
\begin{equation}
\label{eq:eom}
\begin{split}
\sum_{m'}\big[\omega\delta_{n,m'}-H_{n,m'}\big]\mathcal{G}_{m',m}({\boldsymbol{k}},\omega)=\delta_{n,m}\,,
\end{split}
\end{equation}
where $\omega\in[-\Omega/2,\Omega/2]$ and $H_{n,m'}=(H_{\rm sc}-m'\Omega)\delta_{n,m'}+U_{\boldsymbol{k}}\delta_{n+1,m'}+U^{*}_{\boldsymbol{k}}\delta_{n-1,m'}$ is the Floquet Hamiltonian. Here $U_{\boldsymbol{k}}=(1/T)\int_{0}^{T}dt{\rm e}^{i\Omega t} V_{\boldsymbol{k}}(t)= eA_{0}/(2m)(\sigma k_{y}-ik_{x})\tau_{0}$ couples nearest-neighbor sidebands, which involves the emission/absorption of a photon. Finally, the Floquet pair amplitudes $F_{n,m}$  are given by the off-diagonal elements of the  Floquet Green's function $\mathcal{G}_{n,m}$, see  SM \cite{SM}. Note that spin rotation symmetry is preserved such that all pair amplitudes must be spin singlet. 

\begin{figure}[t]
	\centering
	\includegraphics[width=.48\textwidth]{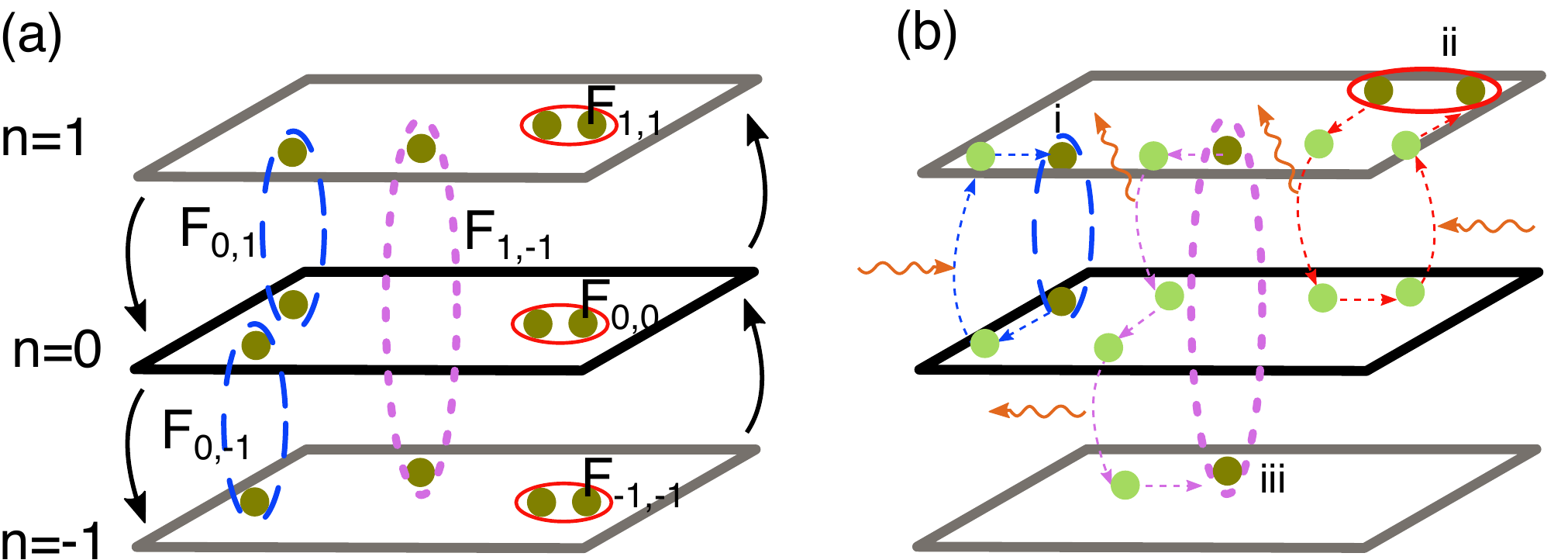}
	\caption{(a) Schematics of the  intra- and inter-sideband Floquet pairs, where black arrows represent the couplings between sidebands. 	(b) Formation of the pairs in (a) involve bare  and dressed processes due to the coupling between sidebands, shown here for  i) $F_{0,1}$ (blue), ii) $F_{1,1}$ (red), and iii) $F_{1,-1}$ (magenta), with absorption/emission of photons (orange wiggle arrows).}
	\label{Fig2}
\end{figure}

The sum over Floquet modes in Eq.\,(\ref{eq:eom}) runs, in principle, to infinity, but since we focus on a finite range of $\omega$, a  truncation of this sum  approximates the exact answer  well \cite{RevModPhys.86.779,rudner2020band,rudner2020floquet}. For instructive purposes, we first restrict our attention to modes  with $n,m\in\{ -1,0,1\}$. 
Even though $\mathcal{G}$ can be directly found from Eq.\,(\ref{eq:eom}),  to visualize its functional dependences,  we consider the limit $U_{\boldsymbol{k}}/\Omega\ll1$ and expand  the Dyson's equation to second order in $U_{\boldsymbol{k}}/\Omega$. Here we show the main findings, while  details are found in the SM \cite{SM}. We find Floquet pair amplitudes of the form $F_{n,m}$ with $n,m\in\{-1,0,1\}$ and for a better analysis we decompose them into their even and odd terms in Floquet indices as $F_{n,m}^{\pm}=(F_{n,m}\pm F_{-m,-n})/2$, obtaining all non-zero amplitudes,
\begin{equation}
\label{FloquetExample}
\begin{split}
F_{0,0}^{+}(\boldsymbol{k},\omega)&\approx \frac{\Delta}{D}+\frac{\Delta|U_{\boldsymbol{k}}|^{2}A_{\omega}^{\Omega}}{D^{2}D_{-1}D_{1}}\,,\\
F_{1,1}^{+}(\boldsymbol{k},\omega)&\approx 
\frac{\Delta[D+\Omega^{2}]}{D_{1}D_{-1}}
+\frac{\Delta |U_{\boldsymbol{k}}|^{2}B_{\omega}^{\Omega}}{2D(D_{-1}D_{1})^{2}}\,,\\
F_{1,1}^{-}({\boldsymbol{k}},\omega)&\approx 
\frac{-2\omega\Omega\Delta}{D_{1}D_{-1}}
+ \frac{2\omega\Omega\Delta |U_{\boldsymbol{k}}|^{2}C_{\omega}^{\Omega}}{D(D_{-1}D_{1})^{2}}\,,\\
F^{+}_{1,-1}(\boldsymbol{k},\omega)&\approx-\frac{\Delta [U_{\boldsymbol{k}}^{*}]^{2}E_{\omega}^{\Omega}}{DD_{1}D_{-1}}\,,\\
F_{0,1}^{+}(\boldsymbol{k},\omega)&\approx-\frac{2\omega\Delta U_{\boldsymbol{k}} }{D_{1}D_{-1}}\,,\\
F_{0,1}^{-}(\boldsymbol{k},\omega)&\approx- \frac{\Omega\Delta U_{\boldsymbol{k}} E_{\omega}^{\Omega}}{D D_{1}D_{-1}}\,,
\end{split}
\end{equation}
where $D(\omega)=\omega^{2}-(\Delta^{2}+\xi_{\boldsymbol{k}}^{2})$, 
and $D_{n}(\omega)=D(\omega+n\Omega)$. Here, $A_{\omega}^{\Omega}$, $B_{\omega}^{\Omega}$, $C_{\omega}^{\Omega}$, $E_{\omega}^{\Omega}$, $D_{1}D_{-1}$ are even functions of  ${\boldsymbol{k}}$ and  $\omega$, whose explicit expressions are not necessary for our discussion but  given in the SM \cite{SM}. The first three expressions in Eqs.\,(\ref{FloquetExample}) describe Cooper pairs that  emerge within each sideband (intra-sideband), while the last three describe pairs between electrons in different sidebands (inter-sideband),  as  depicted in Fig.\,\ref{Fig2}(a). All these pairs contain both bare and dressed processes, due to the coupling between sidebands via $U_{\boldsymbol{k}}$, shown in Fig.\,\ref{Fig2}(b) for three illustrative cases, which involve  absorption/emission of photons  (orange wiggle arrows).

The intra-sideband amplitudes, $F_{0,0}^{+}$ and $F_{1,1}^{\pm}$ in Eqs.\,(\ref{FloquetExample}),  include bare contributions (first term on the right hand side) and corrections proportional to $|U_{\boldsymbol{k}}|^{2}$, that involve transitions between sidebands assisted by two-photon (emission \textit{and} absorption of a photon) processes, see e.g.~process ii) in Fig.\,\ref{Fig2}(b) and SM \cite{SM}. We verify that higher order corrections  always require an even number of photons. The amplitude for $n=m=0$ is purely even in Floquet indices, $F^{+}_{0,0}$, even-$\omega$, and even in $\boldsymbol{k}$, thus belonging to symmetry  class  1 in Table \ref{Table1}.  Interestingly,  we find that $n,m=\pm1$ pair amplitudes develop both even and odd terms in Floquet indices, $F_{1,1}^{\pm}$.  Here $F^{+}_{1,1}$  is even in both frequency and momentum, thus belonging to class 1 in Table \ref{Table1}, while $F^{-}_{1,1}$ instead clearly has an odd-$\omega$ dependence even at zeroth order in $U_{\boldsymbol{k}}$, which is directly controllable by the drive frequency $\Omega$.  We note that $F^{-}_{1,1}$ is also odd in the Floquet indices but even in momentum and thus belongs to symmetry class 8 in Table \ref{Table1}.  These results highlight a key aspect:  the characterization of pair symmetries of periodically-driven superconducting states of matter is unique and intrinsically different from the equilibrium case and can be induced and controlled  by time-periodic fields.

In contrast to the intra-sideband case, the inter-sideband  pair amplitudes in Eqs.\,(\ref{FloquetExample}), always require transitions between sidebands (via $U_{\boldsymbol{k}}$) and thus involve absorption \textit{or} emission of photons.  We distinguish between two types of pair amplitudes, requiring nearest-neighbor (e.g.~$F_{0,1}^{\pm}$ between $n=0$ and $n=1$)  or next-nearest-neighbor transitions (e.g. $F_{1,-1}^{+}$). Here, $F_{0,1}^{\pm}$ is linear in $U_{\boldsymbol{k}}$ and only necessitates a one-photon process as depicted in process i) Fig.\,\ref{Fig2}(b).  However, $F_{1,-1}^{+}$ is proportional to $[U_{\boldsymbol{k}}^{*}]^{2}$ and thus needs two-photon processes of the same kind, either absorption \textit{or} emission of two photons, as depicted in process iii)  Fig.\,\ref{Fig2}(b). Note this two-photon process is still fundamentally different from the intra-sideband amplitudes.  
Here we find that only the  even Floquet index component $F_{0,1}^{+}$ develops an odd-$\omega$ term, which is proportional to $\omega$, instead of $\omega\Omega$ found for intra-sideband terms. Surprisingly, it is also odd in momentum $\boldsymbol{k}$, class 6 in Table \ref{Table1}, and thus of $p$-wave nature, despite the $s$-wave symmetry of the superconductor. This is a consequence of the linear coupling between the light field and momentum. Finally, we find that $F_{0,1}^{-}$ and $F_{1,-1}^{+}$ belong to 
classes 3 and 1, respectively,  in Table \ref{Table1}.

\begin{figure}[t]
	\centering
	\includegraphics[width=.5\textwidth]{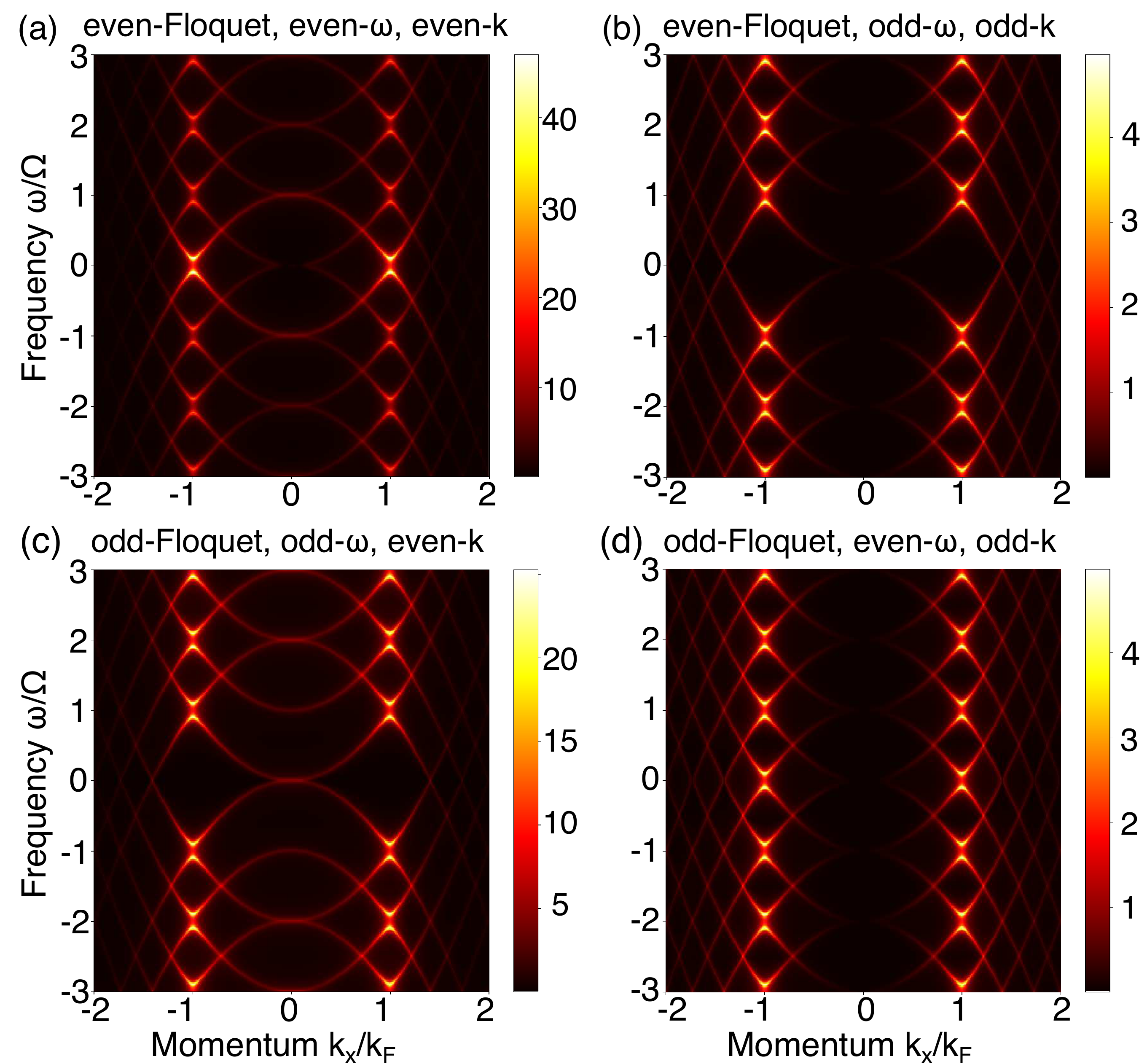}
	\caption{Floquet pair amplitudes in the $\omega$,$k_x$-plane, projected onto the symmetry classes of Table \ref{Table1}: (a) class 1, (b) class 6, (c) class 8, and (d) class 3.   Parameters: $\Delta=\Omega/10$, $eA_0/2m=\Omega/20$, $2m=1$, $N=10$.
	}
	\label{fig:pair_amplitudes}
\end{figure}

After analytically confirming the emergence of Floquet pair amplitudes $F_{n,m}$ within perturbation theory,  we  proceed to compare the relative sizes of each of the symmetry classes by solving Eq.~(\ref{eq:eom}) numerically to infinite order.  For this purpose we truncate the sum over the Floquet indices, such that  $n,m\in[-N,N]$ for some integer cutoff $N$, and numerically obtain $\mathcal{G}$, whose anomalous components then yield $F_{n,m}$.  Finally, we decompose $F_{n,m}$  into the various symmetry classes of Table \ref{Table1} and we find finite amplitudes belonging to classes 1, 3, 6, and 8, shown in Fig.\,\ref{fig:pair_amplitudes} as a function of $\omega$ and $\boldsymbol{k}$. We verify that the overall physical dependencies on $\omega$ and $\boldsymbol{k}$ of these amplitudes are consistent with the second order results given by Eqs. (\ref{FloquetExample}) and that all other symmetry classes are equal to zero, supporting the validity of the perturbation approach used above. We have also checked that  the results of Fig.\,\ref{fig:pair_amplitudes} remain unchanged for larger values of  $N$ and  do not depend sensitively on the choice of model parameters.

The bright areas of the pair amplitudes  in Fig.\,\ref{fig:pair_amplitudes}  extend to higher $\omega$ and  reflect  the fact that each pair amplitude  contains the contribution from many Floquet components. Indeed, in all panels we observe the energy versus momentum dispersion Floquet replicas, a standard feature of  Floquet superconductors.  These Floquet replicas can be measured e.g.~by time- and angle-resolved photoemission spectroscopy \cite{Wang453,mahmood2016selective}. Since the pair amplitudes exist only when the Floquet replicas are present, the experimental observation of these replicas  constitutes an indirect but strong sign of the emergence of Floquet pair amplitudes. Note that the even- (a,d)  and odd-$\omega$ classes (b,c) exhibit high and low intensities, respectively, near  $\omega=0$, thus enabling their distinction. This can be understood by noting that effects at low $\omega$ stem from the lowest Floquet mode $n=0$ \cite{RevModPhys.86.779,rudner2020band,rudner2020floquet}, and that  $F_{0,0}^{+}$ has even-$\omega$ symmetry, see Eqs. (\ref{FloquetExample}). At finite frequencies (e.g.~at $\omega\approx\Omega$)  the odd-$\omega$ amplitudes also acquire large values due to individual odd-$\omega$ contributions from higher Floquet modes, see e.g.~$F_{1,1}^{-}$ in Eqs.~(\ref{FloquetExample}). Moreover, Fig.\,\ref{fig:pair_amplitudes} shows that the even- and odd-$\omega$ amplitudes are of the same order of magnitude, an unusual feature for odd-$\omega$ correlations in conventional superconductors \cite{Balatsky2017}.  

From the analytical and numerical results presented above we see that, circularly polarized light applied to a conventional $s$-wave spin-singlet superconductor gives rise to a very rich variety of Floquet pair amplitudes, which do not emerge in static systems. In particular, we find substantial odd-$\omega$ pairing which only require breaking the continuous time-translation invariance in time-periodic superconductors, unlike previous studies which needed to break additional symmetries \cite{PhysRevB.94.094518, PhysRevB.95.224518}.   These results not only provide a fundamental understanding of Cooper pairs in driven superconductors, but we also anticipate that the Floquet  pair amplitudes can have important consequences in experimental observables \footnote{A detail study on the experimental consequences of these Floquet pair amplitudes will be presented elsewhere.}. For example, it has been shown that odd-$\omega$ pairing can induce a paramagnetic Meissner contribution in multiband systems \cite{PhysRevB.72.140503,PhysRevB.91.144514,PhysRevX.5.041021,PhysRevB.92.224508,PhysRevB.101.180512,PhysRevLett.125.026802}. However, the magnitude of this paramagnetic effect is small due to the small  odd-$\omega$ term and thus hard to distinguish from the diamagnetic even-$\omega$ contribution. The large Floquet odd-$\omega$ pair amplitudes found here would be expected to remedy that situation. Moreover, our findings might also be relevant for other recent advances in superconductors under time-periodic driving fields, such as  Higgs modes in superconductors under radiation \cite{doi:10.1146/annurev-conmatphys-031119-050813,schwarz2020classification,homann2020higgs}, Floquet Majorana fermions \cite{PhysRevB.87.115420,PhysRevB.90.205127}, and time-crystalline superconductors \cite{PhysRevLett.124.096802}, where the emergence of Floquet odd-$\omega$ pairs should be inevitable and could play an important role.

\section{Conclusions}
In conclusion, we have demonstrated that the symmetry classification of superconducting pair amplitudes is significantly broadened in time-periodic superconductors  by virtue of their Floquet bands. In particular, we have  introduced the concept Floquet odd-frequency pair amplitudes with no analog in static systems and showed that they can be induced and controlled by the drive frequency even in fully conventional superconductors. 
The ability to induce and control  superconducting pair amplitudes via time-periodic fields  \cite{doi:10.1146/annurev-conmatphys-031218-013423,Giovannini_2019,rudner2020band,francesconi2020engineering,porta2020topological}  paves the way for Floquet engineering dynamical superconducting states, whose fundamental understanding allow for the design of novel superconducting devices.

\section{Acknowledgments}
We thank A.~Balatsky, D.~Chakraborty, F.~Parhizgar, and S.~Pradhan for useful discussions. We acknowledge financial support from the Swedish Research Council (Vetenskapsr\aa det Grant No.~2018-03488), the European Research Council (ERC) under the European Unions Horizon 2020 research and innovation programme (ERC-2017-StG-757553), the Knut and Alice Wallenberg Foundation through the Wallenberg Academy Fellows program and the EU-COST Action CA-16218 Nanocohybri.

\bibliography{biblio}

\cleardoublepage
\onecolumngrid
\appendix


\section*{Supplementary material}

# Supplemental Material for "Floquet engineering bulk odd-frequency superconducting pairs"

Jorge Cayao,[1] Christopher Triola,[2,1] and Annica M. Black-Schaffer[1]

[1] *Department of Physics and Astronomy, Uppsala University, Box 516, S-751 20 Uppsala, Sweden*
[2] *Los Alamos National Laboratory, Los Alamos, New Mexico 87544, USA*
(Dated: August 13, 2020)

In this supplementary material we give details to further support the results and conclusions of the main text. In particular, we provide additional information on the Floquet Green's functions for time-periodic superconductors and details of the analytical calculation of the Floquet pair amplitudes for a spin singlet $s$-wave superconductor driven by circularly polarized light, presented in Eqs. (4) of the main text.

## FLOQUET GREEN'S FUNCTION IN NAMBU SPACE

In order to find the pair amplitudes we need to calculate the system's Green's function in Nambu space, which reads [1, 2]

$$\hat{\mathcal{G}}(\boldsymbol{k};t_1,t_2) = -i\langle \mathcal{T} \Psi_{\boldsymbol{k}}(t_1)\Psi_{\boldsymbol{k}}^{\dagger}(t_2)\rangle = -i\begin{pmatrix} \langle \mathcal{T} c_{\boldsymbol{k},\uparrow}(t_1)c_{\boldsymbol{k},\uparrow}^{\dagger}(t_2)\rangle & \langle \mathcal{T} c_{\boldsymbol{k},\uparrow}(t_1)c_{-\boldsymbol{k},\downarrow}(t_2)\rangle \\ \langle \mathcal{T} c_{-\boldsymbol{k},\downarrow}^{\dagger}(t_1)c_{\boldsymbol{k},\uparrow}^{\dagger}(t_2)\rangle & \langle \mathcal{T} c_{-\boldsymbol{k},\downarrow}^{\dagger}(t_1)c_{-\boldsymbol{k},\downarrow}(t_2)\rangle \end{pmatrix}. \quad (\text{S1})$$

Here $\Psi_{\boldsymbol{k}} = (c_{\boldsymbol{k},\uparrow}, c_{-\boldsymbol{k},\downarrow}^{\dagger})^{\text{T}}$ is the Nambu spinor, $\mathcal{T}$ is the time-ordering operator, and $c_{\boldsymbol{k},\sigma}(t)$ annihilates a single particle electronic state with spin $\sigma$, momentum $k$, at time $t$. The pair amplitudes are obtained from the anomalous and off-diagonal electron-hole term of Eq. (S1) equation and given by

$$F_{\uparrow,\downarrow}(\boldsymbol{k};t_1,t_2) = -i\langle \mathcal{T} c_{\boldsymbol{k},\uparrow}(t_1)c_{-\boldsymbol{k},\downarrow}(t_2)\rangle. \quad (\text{S2})$$

Note that we are excluding non-zero momentum pairing and also limit ourselves to spin-singlet configurations. Under general grounds, however, the pair amplitude is given by $F_{\sigma_1,\sigma_2}(\boldsymbol{k}_1,t_1;\boldsymbol{k}_2,t_2) = -i\langle \mathcal{T} c_{\sigma_1,\boldsymbol{k}_1}(t_1)c_{\sigma_2,\boldsymbol{k}_2}(t_2)\rangle$, but we restrict ourselves here to the case of zero momentum pairing, as given by Eq. (S2).

Under a time-periodic field $E(t)$, with period $T = 2\pi/\Omega$ and frequency $\Omega$, all system quantities acquire a periodicity in time. In particular, the Green's functions, defined above and dependent on two times $t_1$ and $t_2$, are also periodic in these times, i.e. $\hat{\mathcal{G}}(\boldsymbol{k};t_1,t_2) = \hat{\mathcal{G}}_{\sigma_1,\sigma_2}(\boldsymbol{k};t_1+T,t_2+T)$. The Floquet theorem allows us to Fourier decompose $\hat{\mathcal{G}}$ into harmonics of the fundamental drive frequency $\Omega$ as [3, 4]

$$\hat{\mathcal{G}}(\boldsymbol{k};t_1,t_2) = \sum_{m,n}\int_{-\Omega/2}^{\Omega/2} d\omega\, \mathcal{G}_{n,m}(\boldsymbol{k},\omega)\mathrm{e}^{-i(\omega+n\Omega)t_1+i(\omega+m\Omega)t_2}, \quad (\text{S3})$$

where $\mathcal{G}_{n,m}$ are $2\times 2$ matrices in electron-hole subspace with the off-diagonal terms again giving the pair amplitudes. Note that the Floquet decomposition is standard for two-time Green's functions [3, 4] and, therefore, it can also be applied to the anomalous electron-hole component in Eq. (S3), as performed in the section "Floquet pair amplitudes" of the main text.

To find the pair amplitudes, we calculate the system's Green's function and then extract the pair amplitudes. For this purpose, we write down the equation of motion for $\hat{\mathcal{G}}$ as [3],

$$\sum_{m'}\left[(\omega + m'\Omega - H_{\text{sc}})\delta_{n,m'} - U_{\boldsymbol{k}}\delta_{n+1,m'} - U_{\boldsymbol{k}}^{*}\delta_{n-1,m'}\right]\mathcal{G}_{m',m}(\boldsymbol{k},\omega) = \delta_{n,m}, \quad (\text{S4})$$

where $\omega \in [\Omega/2, \Omega/2]$ and $U_{\boldsymbol{k}} = eA_0/(2m)(\sigma k_y - ik_x)\tau_0$ couples nearest-neighbor Floquet bands or sidebands, with a Floquet index $n$, and involves the emission/absorption of one photon. Note that the elements $\mathcal{G}_{n,m}$ in Eq. (S4) are matrices in Nambu space,

$$\mathcal{G}_{n,m}(\boldsymbol{k},\omega) = \begin{pmatrix} G_{n,m}(\boldsymbol{k},\omega) & F_{n,m}(\boldsymbol{k},\omega) \\ F_{n,m}^{\dagger}(\boldsymbol{k},\omega) & G_{n,m}^{\dagger}(\boldsymbol{k},\omega) \end{pmatrix} \quad (\text{S5})$$

where $G_{n,m}$ and $F_{n,m}$ are the Floquet normal and anomalous Green's function components. The equation of motion given by Eq. (S4) corresponds to Eq. (3) in the main text.

# PERTURBATIVE CALCULATION OF THE FLOQUET PAIR AMPLITUDES

In this section we derive Eqs. (4) in the main text. For a clear presentation of the derivation we divide the section into three subsections.

### Truncation of the Floquet Green's function

The sum over Floquet modes in Eq. (S4) runs, in principle, to infinity, but since we are only interested in a finite range of energies, a finite truncation of this sum approximates the exact answer very well [3–5]. To derive Eqs. (4) in the main text, we restrict our attention to the modes in Eq. (S4) with $n, m \in \{-1, 0, 1\}$. This simple approach nicely captures all the functional dependences of the pair amplitudes and highlights the role of the Floquet bands. For the first three sidebands $n, m \in \{-1, 0, 1\}$, the Floquet Green's function from Eq. (S4) can be written as

$$\mathcal{G}_\mathrm{F}(\boldsymbol{k}, \omega) = \begin{pmatrix} \omega - H_\mathrm{sc} - \hbar\Omega & -U_{\boldsymbol{k}}^* & 0 \\ -U_{\boldsymbol{k}} & \omega - H_\mathrm{sc} & -U_{\boldsymbol{k}}^* \\ 0 & -U_{\boldsymbol{k}} & \omega - H_\mathrm{sc} + \hbar\Omega \end{pmatrix}^{-1} = \begin{pmatrix} g_{-1,-1}^{-1}(\boldsymbol{k},\omega) & -U_{\boldsymbol{k}}^* & 0 \\ -U_{\boldsymbol{k}} & g_{0,0}^{-1}(\boldsymbol{k},\omega) & -U_{\boldsymbol{k}}^* \\ 0 & -U_{\boldsymbol{k}} & g_{1,1}^{-1}(\boldsymbol{k},\omega) \end{pmatrix}^{-1}, \quad (S6)$$

where the momentum $\boldsymbol{k}$ dependence is found in the static Hamiltonian of the superconductor $H_\mathrm{sc}$ and $U_{\boldsymbol{k}}^*$ and $U_{\boldsymbol{k}}$ couple nearest-neighbor sidebands. Moreover, we have

$$\begin{aligned} g_{0,0}(\boldsymbol{k},\omega) &= (\omega - H_\mathrm{sc})^{-1} = \frac{1}{D(\omega)} \begin{pmatrix} \omega + \xi_{\boldsymbol{k}} & \Delta \\ \Delta & \omega - \xi_{\boldsymbol{k}} \end{pmatrix}, \\ g_{-1,-1}(\boldsymbol{k},\omega) &= g_{0,0}(\boldsymbol{k},\omega - \Omega) = \frac{1}{D(\omega - \Omega)} \begin{pmatrix} \omega - \Omega + \xi_{\boldsymbol{k}} & \Delta \\ \Delta & \omega - \Omega - \xi_{\boldsymbol{k}} \end{pmatrix}, \\ g_{1,1}(\boldsymbol{k},\omega) &= g_{0,0}(\boldsymbol{k},\omega + \Omega) = \frac{1}{D(\omega + \Omega)} \begin{pmatrix} \omega + \Omega + \xi_{\boldsymbol{k}} & \Delta \\ \Delta & \omega + \Omega - \xi_{\boldsymbol{k}} \end{pmatrix}, \end{aligned} \quad (S7)$$

which are the bare Green's functions within each sideband $n = 0, -1, 1$, respectively, with

$$D(\omega) = \omega^2 - (\Delta^2 + \xi_{\boldsymbol{k}}^2), \qquad D(\omega \mp \Omega) = D(\omega) + \Omega^2 \mp 2\omega\Omega. \quad (S8)$$

### Dyson's approach for the Floquet Green's function

One possibility for obtaining $\mathcal{G}_\mathrm{F}$ is to directly perform the matrix inversion in Eq. (S6). It is also possible to analytically perform perturbation theory in the coupling between sidebands $U_{\boldsymbol{k}}$, which allow for more insights on the role of the sidebands, couplings, and bare propagators $g_{nn}$. This can be done by projecting the Dyson's equation $G = g + gVG$ on the subspace of $n = 0, -1, 1$, thus obtaining all the elements of the Floquet Green's function, Eq. (S4). Here, $G$ is the dressed Green's function, $g$ represents the bare propagator, and $V$ the coupling between sidebands. Up to second order in $V$ we obtain

$$\mathcal{G}_{nm} \approx \langle n | G | m \rangle \approx \langle n | g | m \rangle + \langle n | gVg | m \rangle + \langle n | gVgVg | m \rangle \quad (S9)$$

where $n, m = 0, \pm 1$, and $|n\rangle$ is the state of the $n$th sideband. Note that $\langle n | g | m \rangle = g_{nm}$ is finite only when $n = m$ because they describe intra-sideband propagators. Also, $V$ couples different sidebands, where $\langle r | V | s \rangle = V_{rs}$ is finite only for $r \neq s$ being nearest neighbor sidebands. Note that transitions between next nearest-neighbor sidebands are not allowed due to the nature of the drive, see Eqs. (S4) and (S6).

With these considerations, we arrive, within second order in the coupling between sidebands, at the following

expressions for the elements of the Floquet Green's functions,

$$\begin{aligned}
\mathcal{G}_{00} &\approx g_{00} + g_{00}V_{01}g_{11}V_{10}g_{00} + g_{00}V_{0-1}g_{-1-1}V_{-10}g_{00} \\
\mathcal{G}_{01} &\approx g_{00}V_{01}g_{11}, \\
\mathcal{G}_{0-1} &\approx g_{00}V_{0-1}g_{-1-1}, \\
\mathcal{G}_{10} &\approx g_{11}V_{10}g_{00}, \\
\mathcal{G}_{-10} &\approx g_{-1-1}V_{-10}g_{00}, \\
\mathcal{G}_{1-1} &\approx g_{11}V_{10}g_{00}V_{0-1}g_{-1-1}, \\
\mathcal{G}_{-11} &\approx g_{-1-1}V_{-10}g_{00}V_{01}g_{11}, \\
\mathcal{G}_{-1-1} &\approx g_{-1-1} + g_{-1-1}V_{-10}g_{00}V_{0-1}g_{-1-1}, \\
\mathcal{G}_{11} &\approx g_{11} + g_{11}V_{10}g_{00}V_{01}g_{11},
\end{aligned} \tag{S10}$$

where the bare propagators $g_{nm}$ are given by Eqs. (S7), while $V_{01} = V_{-10} = -U_{\boldsymbol{k}}$ and $V_{10} = V_{0-1} = -U_{\boldsymbol{k}}^*$, and we have suppressed the dependence on frequency $\omega$ and momentum $\boldsymbol{k}$ for brevity.

In Eqs. (S10) it is clear that the sidebands contribute to each dressed propagator. This approximation is the simplest possible way to incorporate the role of different sidebands. The Greens functions given by Eqs. (S10) have a very physical interpretation. For instance, the intra-sideband Green's function $G_{00}$ includes the original bare term $g_{00}$ but also terms that start at $n = 0$ with $g_{00}$, jump to $n = 1$, propagate in $n = 1$ with $g_{11}$, jump back to $n = 0$, and finally propagate in $n = 0$ with $g_{00}$. The jumps need a finite value of the coupling between sidebands and physically due to the emission or absorption of a photon. Likewise, for the other dressed Green's functions. It is interesting to notice that, in order for the dressed processes to occur, photon assisted processes need to occur, otherwise only bare propagators exist. The structure of Eqs. (S10) allows us to illustrate schematically processes i)-iii) in Fig. 2(b) in the main text.

### Floquet pair amplitudes

By using Eqs. (S7) and (S8) we obtain the elements of the Floquet Green's function Eq. (S6) for the first three sidebands $-1, 0, 1$ within second order in the coupling between sidebands. In what follows, we restrict our attention to the superconducting pair amplitudes $F_{n,m}$ which are obtained from the anomalous electron-hole terms of $\mathcal{G}_{n,m}$. Thus, we get,

$$\begin{aligned}
F_{0,0}(\boldsymbol{k},\omega) &\approx f_{00}(\boldsymbol{k},\omega) + \frac{\Delta |U_{\boldsymbol{k}}|^2 \left[2(\Omega^2 + D)(4\omega^2 - D) - 8(\omega\Omega)^2\right]}{D^2 D_1 D_{-1}}, \\
F_{1,1}(\boldsymbol{k},\omega) &\approx f_{11}(\boldsymbol{k},\omega) + \frac{\Delta |U_{\boldsymbol{k}}|^2}{D D_1^2}\left[4\omega^2 - D + \Omega^2 + 4\omega\Omega\right], \\
F_{-1,-1}(\boldsymbol{k},\omega) &\approx f_{-1-1}(\boldsymbol{k},\omega) + \frac{\Delta |U_{\boldsymbol{k}}|^2}{D D_{-1}^2}\left[4\omega^2 - D + \Omega^2 - 4\omega\Omega\right], \\
F_{1,0}(k,\omega) &\approx -\frac{\Delta U_{\boldsymbol{k}}^*}{D_0 D_1}(2\omega + \Omega), \\
F_{0,1}(\boldsymbol{k},\omega) &\approx -\frac{\Delta U_{\boldsymbol{k}}}{D_0 D_1}(2\omega + \Omega), \\
F_{-1,0}(\boldsymbol{k},\omega) &\approx -\frac{\Delta U_{\boldsymbol{k}}}{D_0 D_{-1}}(2\omega - \Omega), \\
F_{0,-1}(\boldsymbol{k},\omega) &\approx -\frac{\Delta U_{\boldsymbol{k}}^*}{D_0 D_{-1}}(2\omega - \Omega), \\
F_{1,-1}(\boldsymbol{k},\omega) &\approx \frac{\Delta [U_{\boldsymbol{k}}^*]^2}{D D_1 D_{-1}}\left(4\omega^2 - D - \Omega^2\right), \\
F_{-1,1}(\boldsymbol{k},\omega) &\approx \frac{\Delta U_{\boldsymbol{k}}^2}{D D_1 D_{-1}}\left(4\omega^2 - D - \Omega^2\right),
\end{aligned} \tag{S11}$$

where $D$ and $D_{\pm 1} = D(\omega \pm \Omega)$ are given by Eqs. (S8), $D_1 D_{-1} = \Omega^2 [\Omega^2 - 2(\Delta^2 + \xi_k^2 + \omega^2)] + D^2$, and

$$f_{0,0}(\boldsymbol{k},\omega) = \frac{\Delta}{D(\omega)}, \qquad f_{-1,-1}(\boldsymbol{k},\omega) = \frac{\Delta\{[D(\omega) + \Omega^2] + 2\omega\Omega\}}{(D(\omega) + \Omega^2)^2 - (2\Omega\omega)^2}, \qquad f_{1,1}(\boldsymbol{k},\omega) = \frac{\Delta\{[D(\omega) + \Omega^2] - 2\omega\Omega\}}{(D(\omega) + \Omega^2)^2 - (2\Omega\omega)^2}, \quad (S12)$$

correspond to the bare intra-sideband pair amplitudes obtained from the off-diagonal components of the bare Green's functions in Eqs. (S7).

The expressions given by Eqs. (S11) represent Floquet Cooper pairs in a conventional spin-singlet $s$-wave superconductor driven by circularly polarized light. Floquet pair amplitudes can thus emerge as intra-sideband, such as $F_{0,0}$, $F_{1,1}$, and $F_{-1,-1}$, or as inter-sideband, such as $F_{0,1}$, $F_{0,-1}$, $F_{1,0}$, $F_{-1,0}$, $F_{1,-1}$, and $F_{-1,1}$, which all involve emission or/and absorption of photons due to transitions between sidebands.

Finally we analyze the symmetries of the Floquet pair amplitudes given by Eqs. (S11) to determine the nature of Floquet Cooper pairs. For this purpose we write the even and odd projection in Floquet indices,

$$F_{n,m}^{\pm}(\boldsymbol{k},\omega) = \frac{F_{n,m}(\boldsymbol{k},\omega) \pm F_{-m,-n}(\boldsymbol{k},\omega)}{2} \tag{S13}$$

where we have used the relation $(n,m) \to (-m,-n)$ for the pair amplitudes under the exchange of Floquet indices, presented in Eq. (2) of the main text. Then, we obtain

$$\begin{aligned}
F_{0,0}^{+}(\boldsymbol{k},\omega) &\approx f_{0,0}^{+}(\boldsymbol{k},\omega) + \frac{\Delta |U_{\boldsymbol{k}}|^2 A_\omega^\Omega}{D^2 D_{-1} D_1}, \\
F_{0,0}^{-}(\boldsymbol{k},\omega) &= 0, \\
F_{1,1}^{+}(\boldsymbol{k},\omega) &\approx f_{1,1}^{+}(\boldsymbol{k},\omega) + \frac{\Delta |U_{\boldsymbol{k}}|^2 B_\omega^\Omega}{2D(D_{-1} D_1)^2} \\
F_{1,1}^{-}(\boldsymbol{k},\omega) &\approx f_{1,1}^{-}(\boldsymbol{k},\omega) + \frac{2\omega\Omega\Delta |U_{\boldsymbol{k}}|^2 C_\omega^\Omega}{D(D_{-1} D_1)^2} \\
F_{1,-1}^{+}(\boldsymbol{k},\omega) &\approx -\frac{\Delta [U_{\boldsymbol{k}}^*]^2 E_\omega^\Omega}{D D_1 D_{-1}}, \\
F_{1,-1}^{-}(\boldsymbol{k},\omega) &= 0, \\
F_{0,1}^{+}(\boldsymbol{k},\omega) &\approx -\frac{2\omega \Delta U_{\boldsymbol{k}}}{D_1 D_{-1}}, \\
F_{0,1}^{-}(\boldsymbol{k},\omega) &\approx -\frac{\Omega \Delta U_{\boldsymbol{k}} E_\omega^\Omega}{D D_1 D_{-1}},
\end{aligned} \tag{S14}$$

where

$$f_{0,0}^{+}(\boldsymbol{k},\omega) = \frac{\Delta}{D}, \quad f_{1,1}^{+}(\boldsymbol{k},\omega) = \frac{\Delta[D + \Omega^2]}{D_1 D_{-1}}, \quad f_{1,1}^{-}(\boldsymbol{k},\omega) = \frac{-2\omega\Omega\Delta}{D_1 D_{-1}}, \tag{S15}$$

are the bare components of the intra-sideband pair amplitudes and

$$\begin{aligned}
A_\omega^\Omega &= 2(\Omega^2 + D)(4\omega^2 - D) - 8(\omega\Omega)^2, \\
B_\omega^\Omega &= 16\omega^2\Omega^2(2\omega^2 - \Omega^2) - 2(D^2 - \Omega^4) + 8\omega^2 D(D - 3\Omega^2), \\
C_\omega^\Omega &= 4D(D + \Omega^2 - 2\omega^2), \\
E_\omega^\Omega &= D + \Omega^2 - 4\omega^2
\end{aligned} \tag{S16}$$

are all even functions in $\boldsymbol{k}$ and $\omega$. Note that $F_{0,0}^{-}$ and $F_{1,-1}^{-}$ are zero, as expected from Eqs. (S13). The expressions given in Eqs. (S14) correspond to Eqs. (4) in the main text. Notice that at first sight one might think that Eqs. (S14) do not include quite all the pair amplitudes listed in Eqs. (S11). However, this is still all terms because in Eqs. (S13) we have written only components that are explicitly even and odd in Floquet indices amplitudes. For instance, the amplitude $F_{-1,-1}$ does not appear in Eqs. (S14) but it still contributes to $F_{1,1}^{\pm}$, as follows directly from Eqs. (S13). The same argument applies to the rest of pair amplitudes that do not either appear in Eqs. (S14).

---

[2] G. D. Mahan, *Many-particle physics* (Springer Science & Business Media, 2013).
[3] H. Aoki, N. Tsuji, M. Eckstein, M. Kollar, T. Oka, and P. Werner, Rev. Mod. Phys. **86**, 779 (2014).
[4] M. S. Rudner and N. H. Lindner, arXiv:2003.08252 (2020).
[5] M. S. Rudner and N. H. Lindner, Nat. Rev. Phys. **2**, 229 (2020).



\end{document}